\documentstyle[twoside,fleqn,npb,epsfig]{article}
\newcommand{\AmS}{{\protect\the\textfont2
  A\kern-.1667em\lower.5ex\hbox{M}\kern-.125emS}}

\title{
\begin{flushright}\normalsize
  \vspace{-4cm}
  CERN-TH/99-143 \\
  May 1999
\end{flushright}
High-$t$ Diffraction
\thanks{Talk presented at 7th International
  Workshop on Deep Inelastic Scattering, Zeuthen, Germany, April
  1999.}
}

\author{J.~R.~Forshaw\address{TH-Division, CERN, 1211 Geneva 23, Switzerland}%
\thanks{On leave of absence from the Department of Physics and Astronomy,
        University of Manchester, Brunswick Street, Manchester M13 9PL. UK}}

\begin{document}

\thispagestyle{empty}

\begin{abstract}
The study of those rapidity gap processes where a large momentum is transferred
across the rapidity gap provides an ideal opportunity to understand the
gap producing mechanism wholly within the framework of QCD perturbation
theory. The current theoretical and experimental status of this `high-$t$ 
diffraction' is reviewed.
\end{abstract}

\maketitle

\section{Introduction}
High-$t$ diffraction is the scattering of two particles $A$ and $B$ into a
final state which is made up wholly of two systems $X$ and $Y$ which are
far apart in rapidity and between which there is a large relative transverse
momentum, $P_t$. The square of the exchanged four-momentum is 
$t \approx -P_t^2$. High $t$ means that $-t \gg \Lambda_{{\rm QCD}}^2$. By
studying this class of events we can hope to gain valuable insight into
the Regge limit of strong interactions using the methods of perturbative QCD.
It has been shown that large $t$ is a very effective way of squeezing the
rapidity gap producing mechanism to short distances regardless of
the sizes of the external particles \cite{FS}.

Perturbative QCD in the Regge limit leads to the production of rapidity gaps
via the exchange of a pair of interacting reggeised gluons in an overall 
colour singlet configuration (a reggeised gluon can be thought of as a colour 
octet compound state of any number of ordinary gluons). To leading 
logarithmic accuracy this colour singlet exchange is described by the BFKL 
equation \cite{BL} and predicts a strong rise in the cross-section as the 
rapidity gap increases. It should be remembered that a characteristic of all 
leading logarithmic BFKL calculations is a large uncertainty in the overall 
normalisation of cross-sections. Next-to-leading logarithmic corrections to 
the BFKL equation have not yet been computed for $t \ne 0$. 

\section{Gaps between jets}
One way of selecting high-$t$ events is to look for events which contain
one or more jets in system $X$ and one or more in system $Y$ \cite{Bj,MT}. 
The leading contribution
to this process comes from the elastic scattering of partons via colour
singlet exchange; the outgoing partons hadronise to produce the jets which
are seen. Experiments at FNAL and HERA have reported their first
results on the fraction of all dijet events which contain a gap in rapidity
\cite{zeus,H1,DZero,CDF}.
For rapidities greater than about 3 units between the jet centres, all see a 
clear excess of rapidity gap events over the number expected in the absence
of a strongly interacting colour singlet exchange. It is worth recalling that 
a large excess of gap events at high-$t$ cannot be explained by traditional 
soft pomeron exchange since such a contribution will have died away long 
before due to shrinkage. 

\begin{figure} 
\centerline{\epsfig{file=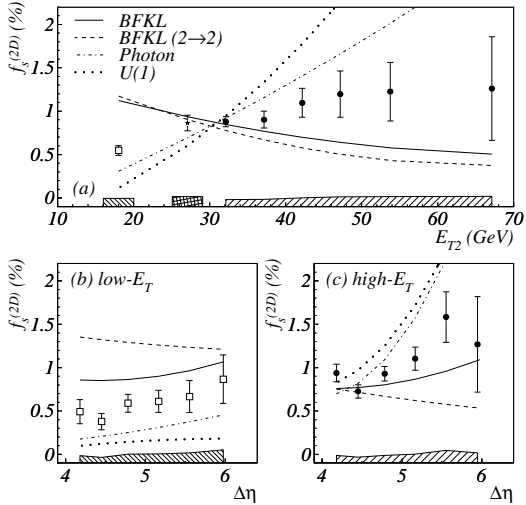,height=6.7cm,angle=0}}
\vspace*{-1cm}
\caption{D0 data compared to different Monte Carlo models. The solid
line is the BFKL prediction produced by HERWIG. Plot from \cite{DZero}.}
\label{fig:dzero}
\vspace*{-0.6cm}
\end{figure}

Unfortunately, spectator interactions can spoil the gap. This physics is
poorly understood and is often modelled by an overall `gap survival' factor.
Assuming the gap survival factor depends only upon the centre-of-mass energy
of the colliding beams, the gap fraction data are mostly in agreement with the 
leading order BFKL predictions with $\alpha_s \approx 0.2$ (it doesn't make 
sense to run the coupling in the leading order BFKL formalism). But
the data still have rather large errors and it is too early to draw 
definitive conclusions. Notwithstanding this, there appears to be a
problem for BFKL: the $E_T$ dependence of the D0 gap fraction shows a
rise with increasing $E_T$ whereas the BFKL prediction is flat or falling, 
see Fig.\ref{fig:dzero}. A few comments are in order. 
Firstly, the solid line on Fig.\ref{fig:dzero}
falls as $E_{T2}$ increases as a consequence of the running of the QCD
coupling (the gap fraction goes like $\sim \alpha_s^4/\alpha_s^2$).
Fixing the coupling, as is strictly proper in the leading logarithmic BFKL
approach, leads to an essentially flat $E_{T2}$ distribution. As an aside,
it might be of interest to note that a fixed coupling was needed in order
to explain the high-$t$ data on high energy $p \bar{p}$ elastic scattering 
\cite{DL}. Secondly, it is to be noted that all jets appearing in
Fig.\ref{fig:dzero} are corrected for an underlying event using minimum 
bias data. Typically, this means subtracting around 1 GeV from the jets.
This is true even for the jets which make up the numerator of the 
gap fraction which, by definition, are produced in an environment free of
any underlying event. This correction is not made in generating the 
theoretical prediction (solid line) which, roughly speaking, means that
the theory curve ought to be multiplied by $[E_{T2}/(E_{T2}+1 {\rm GeV})]^4$ 
(since both the numerator and denominator fall as $1/t^2$). For the lowest
bin at $E_{T2} = 18$ GeV this correction factor is $0.8$. The upshot is that 
theory is probably consistent with a flat $E_{T2}$ spectrum at large
$E_{T2}$ falling to $80\%$ of this value in the lowest $E_{T2}$ bin, and this
is not inconsistent with the D0 data \cite{CFL}. 

The gaps between jets process has a number of drawbacks: the rapidity reach
is seriously compromised by the need to see the two jets; there is the
usual theoretical uncertainty in going from partons to jets; there is the
cloudy issue of gap survival to deal with. One way of improving the
situation is to look at the more inclusive double dissociation sample
\cite{BC}. 

\section{Gluon reggeisation}
As mentioned above, gluons reggeise in QCD. The effect of their reggeisation 
can be studied directly in present day experiments by measuring a process
first suggested in \cite{MT}. The process is 
$h_1 + h_2 \to j_1 + j_2 + X$ where $h_1$ and $h_2$ are the incoming
hadrons, $j_1$ and $j_2$ are the outgoing jets which are far apart in
rapidity and $X$ is the rest of the final state subject to the constraint
that, in the region between the two jets, there be no jets with transverse
momentum larger than some value, $\mu$. The $\mu$ dependence of the ratio of 
these events to all dijet events is interesting since it provides a direct
test of gluon reggeisation. The ratio goes (for $\mu \ll P_t$) like
$$ {\rm e}^{2 \Delta \eta (\alpha_g(P_t)-1)} $$
where $\alpha_g(P_t) = 1 - (N_c/\pi) \alpha_s \ln (P_t/\mu).$ 

Another interesting consequence of gluon reggeisation is that, at large $t$,  
it indicates that fixed order perturbation theory can lead one to infer the
wrong physics. To see this, one needs to realise that the first
correction to two-gluon exchange contains a piece like 
$$\Delta = \alpha_s \ln\left( \frac{{\bf k}^2 ({\bf k-q})^2}{t^2}\right)
\ln s $$
where ${\bf k}$ and ${\bf k-q}$ are the transverse momenta of the two
internal gluons (to get to a cross-section requires an integration over
these momenta with a weighting function dependent upon the system to which
they couple). Such a contribution invites one to conclude that the
most important configuration arises when one gluon carries all the momentum
transfer (${\bf q}^2 = -t$) and the other carries none. However this is
not the case, to order $\alpha_s^2$ the contribution goes like
$\Delta^2/2$ and more generally the series exponentiates to $e^\Delta$ which
strongly suppresses the asymmetric configurations. Consequently, the dominant 
configurations are the symmetric ones where the exchanged gluons share 
the momentum transfer equally \cite{FR}.   

\section{Exclusive vector meson and photon production}
At HERA, focussing on the more exclusive subsample of high-$t$ diffractive 
events in which the photon dissociates to either a photon or a vector meson
and nothing else allows one to neatly sidestep the issue of gap survival.
The detection of vector mesons and photons is also clean enough to allow the 
$t$ distribution (here $t$ is essentially determined by the transverse momentum
of the vector particle) to be measured down to small values. H1
has measured the $t$ distribution for $J/\Psi$ production all the way from
$-t \approx 0$ out to $-t \approx 10$ GeV$^2$. The data agree with the
leading order BFKL calculation with $\alpha_s = 0.2$ \cite{West}. In
addition, ZEUS has investigated $\rho$ mesons produced at high $t$ \cite{JC}. 
The ability of the experiments to determine the kinematics without needing to
observe the proton dissociation system is the crucial factor which allows
the HERA experiments to study events with rapidity gaps as large as 6 units.
Gaps of this size are certainly well into the region of theoretical interest. 
Leading logarithmic level calculations now exist for both vector meson
production \cite{FR,VM} and photon production \cite{photon}.
As well as teaching us about colour singlet exchange, high-$t$ vector 
meson production can also help us understand the vector meson production
mechanism. In this respect, ratios of vector meson production will be
particularly useful. 

Over the coming years HERA promises to produce 
high quality data on light and heavy vector meson production, and on photon 
production (where there is no uncertainty due to wavefunction effects).
These data, in combination with the data on the more inclusive processes
discussed above, will set the benchmark against which we can test our ever
evolving understanding of QCD in the high energy domain.  

\vspace*{-0.3cm}

\section*{Acknowledgment}
\vspace*{-0.3cm}
This work was supported by the EU Fourth Framework Programme 
`Training and Mobility of Researchers', Network `Quantum
Chromodynamics and the Deep Structure of Elementary Particles',
contract FMRX-CT98-0194 (DG 12-MIHT).


\vspace*{-0.2cm}

\begin{thebibliography}{9}
\bibitem{FS} J.R.~Forshaw and P.J.~Sutton, Euro. Phys. J. C14 (1998) 285.

\bibitem{BL} I.~Balitsky and L.N.~Lipatov, Sov. J. Nucl. Phys. 28 (1978) 822. 

\bibitem{Bj} J.D.~Bjorken, Phys. Rev. D47 (1992) 101.

\bibitem{MT} A.~H.~Mueller and W.-K.~Tang, Phys. Lett. B284 (1992) 123.

\bibitem{zeus} M.~Derrick et al (ZEUS collaboration), Phys. Lett. B369 (1996) 
55.

\bibitem{H1} H1 Collaboration, ``Rapidity Gaps between Jets in 
Photoproduction at HERA'', contribution to the International Europhysics
Conference on High Energy Physics, August 1997, Jerusalem, Israel. 

\bibitem{DZero} S.~Abachi et al (D0 collaboration), Phys. Rev. Lett. 72 (1994) 
2332; Phys. Rev. Lett. 76 (1996) 734; B.~Abbott et al (D0 collaboration), 
Phys. Lett. B440 (1998) 189.

\bibitem{CDF} F.~Abe et al (CDF collaboration), Phys. Rev. Lett. 74 (1995) 
855; Phys. Rev. Lett 80 (1998) 1156; Phys. Rev. Lett. 81 (1998) 5278.

\bibitem{DL} A.~Donnachie and P.~V.~Landshoff, Z. Phys. C2 (1979) 55, 
erratum-ibid C2 (1979) 372; Phys. Lett. B387 (1996) 637. 

\bibitem{CFL} B.~E.~Cox, J.~R.~Forshaw and L.~L\"onnblad, in preparation.

\bibitem{BC} B.~E.~Cox: ``Double Diffraction Dissociation at large $|t|$ 
from H1'', contribution to the DIS99 Workshop, Zeuthen, Germany, 1999; 
B.~E.~Cox and J.~R.~Forshaw, Phys. Lett. B434 (1998) 133. 

\bibitem{FR} J.~R.~Forshaw and M.~G.~Ryskin, Z. Phys. C68 (1995) 137.

\bibitem{West} H1 Collaboration: ``Production of $J/\psi$ Mesons with 
large $|t|$ at HERA'', contribution to the International Europhysics
Conference on High Energy Physics, Jerusalem, Israel, 1997. 

\bibitem{JC} J.~Crittenden: ``Recent Results from Decay-Angle Analyses 
of $\rho^0$ Photoproduction at High Momentum Transfer from ZEUS'',
contribution to the DIS99 Workshop, Zeuthen, Germany, 1999. 

\bibitem{VM} J.~Bartels, J.~R.~Forshaw, H.~Lotter and M.~W\"usthoff, 
Phys. Lett. B375 (1996) 301; D.Yu.~Ivanov, Phys. Rev. D53 (1996) 3564; 
I.F.~Ginzburg and D.Yu.~Ivanov, Phys. Rev. D54 (1996) 5523; 
D.Yu.~Ivanov and R.~Kirschner, Phys. Rev. D58 (1998) 114026.

\bibitem{photon} D.~Yu.~Ivanov and M.~W\"usthoff (1999), 
Eur. Phys. J. C8 (1999) 107; 
N.~G.~Evanson and J.~R.~Forshaw, hep-ph/9902481, to appear in Phys. Rev. D.
\end{thebibliography}
\end{document}